\documentclass{article}
\usepackage{spconf, amsmath, graphicx , amssymb}
\usepackage{cite, url}

\usepackage{amsthm}
\usepackage{algorithm}
\usepackage[noend]{algpseudocode}
\usepackage{subcaption}
\usepackage{mwe}

\usepackage{fancyhdr}

\usepackage{amsfonts}
\usepackage{array}
\usepackage{textcomp}
\usepackage{stfloats}
\usepackage{verbatim}
\def\BibTeX{{\rm B\kern-.05em{\sc i\kern-.025em b}\kern-.08em
    T\kern-.1667em\lower.7ex\hbox{E}\kern-.125emX}}
\usepackage{balance}
\usepackage{xcolor}

\usepackage{caption}
\usepackage[labelformat=simple]{subcaption}

\usepackage{tikz}
\usetikzlibrary{tikzmark,decorations.pathreplacing}
\usepackage{rotating}

\usepackage{bm}

\newtheorem*{proof*}{Proof}
\newtheorem*{remark*}{Remark}

\algnewcommand\algorithmicforeach{\textbf{for each}}
\algdef{S}[FOR]{ForEach}[1]{\algorithmicforeach\ #1\ \algorithmicdo}

\title{JOINT DOA ESTIMATION AND DISTORTED SENSOR DETECTION UNDER ENTANGLED LOW-RANK AND ROW-SPARSE CONSTRAINTS}
\name{Huiping Huang$^{\star}$, Tianjian Zhang$^{\dag}$, Feng Yin$^{\dag}$, Bin Liao$^{\S}$, Henk Wymeersch$^{\star}$ \vspace{-2mm} 
\thanks{This paper is partially supported by the Vinnova B5GPOS Project under Grant 2022-01640. Huiping Huang and Tianjian Zhang contribute equally. The authors thank Prof. Hing Cheung So at City University of Hong Kong, Prof. Qi Liu at South China University of Technology, Prof. Liang Yu at Shanghai Jiao Tong University, Mr. Chenyu Zhang and Mr. Lin Chen at Harbin Engineering University, and Mr. Qixin Guo at Tongji University, for the discussions on low-rank and sparse decomposition problems.}}
\address{$^{\star}$Chalmers University of Technology, Gothenburg, Sweden  \\
$^{\dag}$SSE, The Chinese University of Hong Kong, Shenzhen, China  \\
$^{\S}$Shenzhen University, Shenzhen, China \vspace{-2mm} }

\begin{document}
\ninept
\maketitle
%

\begin{abstract}
The problem of joint direction-of-arrival estimation and distorted sensor detection has received a lot of attention in recent decades. Most state-of-the-art work formulated such a problem via low-rank and row-sparse decomposition, where the low-rank and row-sparse components were treated in an isolated manner. Such a formulation results in a performance loss. Differently, in this paper, we entangle the low-rank and row-sparse components by exploring their inherent connection. Furthermore, we take into account the maximal distortion level of the sensors. An alternating optimization scheme is proposed to solve the low-rank component and the sparse component, where a closed-form solution is derived for the low-rank component and a quadratic programming is developed for the sparse component. Numerical results exhibit the effectiveness and superiority of the proposed method.
\end{abstract}

\section{Introduction}
\label{introduction}
Direction-of-arrival (DOA) estimation is a crucial topic in array signal processing, with applications in radar, sonar, wireless communications, source localization, and so on \cite{Krim1996, VanTrees2002, Viberg2014, Wymeersch2017, He2023}. Traditional DOA estimation methods assume an ideal sensor array, but real arrays often suffer from distortions due to various factors such as sensor malfunctions and environmental effects. The traditional methods degrade severely when the sensor array encounters gain and phase uncertainty, which is one of the most common sensor distortions.

A significant body of research is dedicated to addressing distorted or nonfunctional sensors \cite{Yeo1999, Vigneshwaran2007, Oliveri2012, Zhu2015, Wang2017Jul, Liu2019June, Pesavento2002, Liao2012, Stoica2001, Ozturk2023}. In \cite{Yeo1999}, a genetic algorithm was employed for array failure correction. An alternative approach is developed in \cite{Vigneshwaran2007}, where a minimal resource allocation network is utilized for DOA estimation during sensor array failure. Notably, this method demands training without failed sensors. In \cite{Oliveri2012}, a Bayesian compressive sensing technique is introduced, but it mandates a noise-free array as a reference. Methods involving difference co-array are explored in \cite{Zhu2015, Wang2017Jul, Liu2019June}. In particular, \cite{Zhu2015} proposed concept where positions corresponding to failed sensors are replaced by virtual sensors, mitigating the impact of sensor failure. However, this approach does not hold when failed sensors occupy the array's first or last positions, or when malfunctions occur symmetrically, leading to gaps in the difference co-array. On the other hand, \cite{Wang2017Jul} and \cite{Liu2019June} constrain arrays to specific sparse structures, such as co-prime and nested arrays. Another line of study requires pre-calibrated sensors, well-documented over decades, see e.g., \cite{Pesavento2002, Liao2012}. These methods rely on some calibrated sensor, which are time- and energy-intensive.

More recently, low-rank and row-sparse decomposition framework is proposed to solve the joint DOA estimation and distorted sensor detection problem by using the iteratively reweighted least squares (IRLS) \cite{Huang2023}. The limitation of this method lies in that it deals with the low-rank and row-sparse matrices separately, and however these two matrices are inherently connected. In this paper, we investigate such an inherent connection, and moreover, we consider the maximal distortion level of the sensors. A novel algorithm is proposed to solve the low-rank and sparse components, which is verified to have better performance than the IRLS method developed in \cite{Huang2023}.

\section{System Model and Problem Statement}
\label{SignalModel}
\begin{figure}[t]
\centering
{\includegraphics[width=0.48\textwidth]{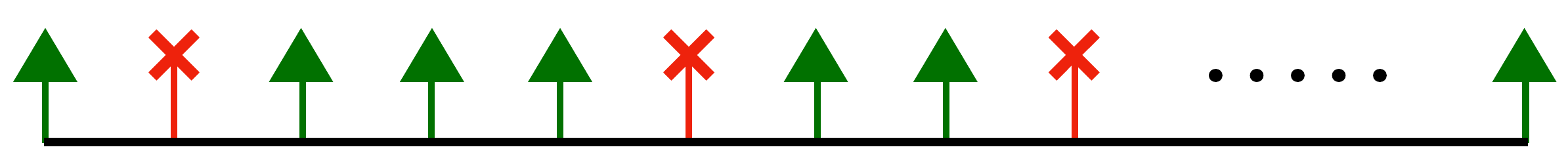}} \vspace{-4mm}
\caption{Illustration of a sparsely distorted linear array, where the red crosses denote distorted sensors.}
\label{SCA}  \vspace{-3mm}
\end{figure}

Suppose that a uniform linear array (ULA) of $M$ sensors receives $K$ far-field narrowband signals from directions ${\bm \theta} = [\theta_{1}, \theta_{2}, \cdots, \theta_{K}]^{\text{T}}$. The antenna array of interest is assumed to be randomly and \textit{sparsely} distorted by sensor gain and phase uncertainty (the number of distorted sensors is smaller than $M$). Further, we assume that the number of distorted sensors and their positions are unknown. Fig. \ref{SCA} illustrates the array model, where the green triangles stand for undistorted sensors and the red crosses refer to distorted ones. The red crosses appear randomly and sparsely within the whole linear array.

The array observation can be written as \cite{Chen2020}:  \vspace{-1mm}
\begin{align*}
{\bf y}(t) = \breve{\bf \Gamma}{\bf A}{\bf s}(t) + {\bf n}(t) = ({\bf I} + {\bf \Gamma}){\bf A}{\bf s}(t) + {\bf n}(t), 
\end{align*}  
where $t = 1, 2, \cdots, T$ denotes the time index, $T$ is the total number of available snapshots, ${\bf s}(t) \in \mathbb{C}^{K}$ and ${\bf n}(t) \in \mathbb{C}^{M}$ are signal and noise vectors, respectively. The steering matrix ${\bf A} = [{\bf a}(\theta_{1}), {\bf a}(\theta_{2}), \cdots, {\bf a}(\theta_{K})] \in \mathbb{C}^{M \times K}$ has steering vectors as columns, where the steering vector ${\bf a}(\theta_{k})$ is a function of $\theta_{k}$, for $k = 1, 2, \cdots, K$. In addition, $\breve{\bf \Gamma} \triangleq {\bf I} + {\bf \Gamma}$ indicates the electronic sensor status (either perfect or distorted), where ${\bf I}$ is the $M \times M$ identity matrix, and ${\bm \Gamma}$ is a diagonal matrix with its main diagonal, ${\bm \gamma} = [\gamma_{1}, \gamma_{2}, \cdots, \gamma_{M}]^{\text{T}}$, being a sparse vector, i.e., ${\bm \Gamma} = \text{diag}\{{\bm \gamma}\}$. Specifically, for $m = 1, 2, \cdots, M$,  \vspace{-1mm}
\begin{align*}
\gamma_{m} \left\{ \begin{array}{l}
			\!\! = 0, ~ \text{if the $m$-th sensor is perfect,} \\
			\!\! \neq 0, ~ \text{if the $m$-th sensor is distorted.}
\end{array} \right.
\end{align*}
The non-zero $\gamma_{m} \in \mathbb{C}$ denotes the sensor gain and phase error. We assume that the sensor error is within some level, that is, $|\mathrm{Re}(\gamma_{m})| \leq \gamma_{\text{max}}$ and $|\mathrm{Im}(\gamma_{m})| \leq \gamma_{\text{max}}$, where $\mathrm{Re}(\cdot)$ and $\mathrm{Im}(\cdot)$ denote the real and imaginary parts of a complex variable, respectively, and $\gamma_{\text{max}}$ is the maximal sensor distortion error (for both real and imaginary parts) and it is known based on prior information.

Collecting all the available $T$ ($T > M$ in general) snapshots into a matrix, we have  \vspace{-1mm}
\begin{align}
\label{datamodel_matrix}
{\bf Y} = ({\bf I} + {\bm \Gamma}){\bf A}{\bf S} + {\bf N},
\end{align}
where ${\bf Y} = [{\bf y}(1), {\bf y}(2), \cdots, {\bf y}(T)] \in \mathbb{C}^{M \times T}$ is the measurement matrix, ${\bf S} = [{\bf s}(1), {\bf s}(2), \cdots, {\bf s}(T)] \in \mathbb{C}^{K \times T}$ is the signal matrix, and ${\bf N} = [{\bf n}(1), {\bf n}(2), \cdots, {\bf n}(T)] \in \mathbb{C}^{M \times T}$ is the noise matrix. Given the array measurements ${\bf Y}$, our task is to jointly estimate the incoming directions of all emitting signals and detect the distorted sensors within the array. Note that the number of distorted sensors is small, but unknown, and their positions are unknown as well.

\section{Proposed Method}
\label{ProposedMethod}
By defining ${\bf Z} \triangleq {\bf A}{\bf S}$ and using ${\bm \Gamma} = \text{diag}\{{\bm \gamma}\}$, (\ref{datamodel_matrix}) becomes:
\begin{align}
\label{datamodel_ZV}
{\bf Y} = {\bf Z} + \text{diag}\{{\bm \gamma}\}{\bf Z} + {\bf N},
\end{align}
where ${\bf Z} \in \mathbb{C}^{M \times T}$ is a low-rank matrix of rank $K$ (in general $K < \min\{M, T\}$). The problem to be solved is formulated as  \vspace{-1mm}
\begin{subequations}
\label{problem_rankl0}
 \begin{align}
    \min_{{\bf Z}, {\bm \gamma}} ~ & \left\{ \!\!\! 
    \begin{array}{l}
    \frac{1}{2}\|{\bf Y} \! - \! {\bf Z} \! - \! \text{diag}\{{\bm \gamma}\}{\bf Z}\|_{\text{F}}^{2} + \lambda_{1}\text{Rank}\{{\bf Z}\} \vspace{1.5mm} \\ 
    + \lambda_{2}(\|\mathrm{Re}({\bm \gamma})\|_{0} + \|\mathrm{Im}({\bm \gamma})\|_{0}),
    \end{array} \right.  \\
    \mathrm{s.t.} ~&~ |\mathrm{Re}(\gamma_{m})| \! \leq \! \gamma_{\text{max}}, |\mathrm{Im}(\gamma_{m})| \! \leq \! \gamma_{\text{max}}, ~ m = 1, 2, \! \cdots \!, M,
 \end{align}    
\end{subequations}
where $\lambda_{1} > 0$ and $\lambda_{2} > 0$ are two tuning parameters, $\|\cdot\|_{\text{F}}$, $\text{Rank}\{\cdot\}$, and $\|\cdot\|_{0}$ are the Frobenius norm, rank function, and $\ell_{0}$ norm, respectively. Note that both $\mathrm{Re}({\bm \gamma})$ and $\mathrm{Im}({\bm \gamma})$ have sparsity structure.

\subsection{Convex relaxation}
The rank function and the $\ell_{0}$ norm are non-convex, which are usually replaced by their convex surrogates, i.e., the nuclear norm and $\ell_{1}$ norm, respectively. Therefore, Problem (\ref{problem_rankl0}) becomes:  \vspace{-1mm}
\begin{subequations}
\label{problem_nuclearl1}
 \begin{align}
    \min_{{\bf Z}, {\bm \gamma}} ~ & \left\{ \!\!\! 
    \begin{array}{l}
    \frac{1}{2}\|{\bf Y} \! - \! {\bf Z} \! - \! \text{diag}\{{\bm \gamma}\}{\bf Z}\|_{\text{F}}^{2} + \lambda_{1}\| {\bf Z} \|_{*} \vspace{1.5mm} \\ 
    + \lambda_{2}(\|\mathrm{Re}({\bm \gamma})\|_{1} + \|\mathrm{Im}({\bm \gamma})\|_{1}),
    \end{array} \right.  \\
    \mathrm{s.t.} ~&~ |\mathrm{Re}(\gamma_{m})| \leq \gamma_{\text{max}}, |\mathrm{Im}(\gamma_{m})| \leq \gamma_{\text{max}}, ~ \forall m,  
 \end{align}  
\end{subequations}
where $\|\cdot\|_{*}$ and $\|\cdot\|_{1}$ stand for the nuclear norm and $\ell_{1}$ norm, respectively. In order to handle the non-smoothness of the nuclear norm, we introduce a smoothing parameter $\mu$ \cite{Lu2015}. That is, Problem (\ref{problem_nuclearl1}) is transferred to  \vspace{-1mm}
\begin{subequations}
\label{problem_nuclearl1_new}
 \begin{align}
    \min_{{\bf Z}, {\bm \gamma}} ~ & \left\{ \!\!\! 
    \begin{array}{l}
    \frac{1}{2}\|{\bf Y} \! - \! {\bf Z} \! - \! \text{diag}\{{\bm \gamma}\}{\bf Z}\|_{\text{F}}^{2} + \lambda_{1}\| [{\bf Z} , \mu {\bf I}] \|_{*} \vspace{1.5mm} \\ 
    + \lambda_{2}(\|\mathrm{Re}({\bm \gamma})\|_{1} + \|\mathrm{Im}({\bm \gamma})\|_{1}),
    \end{array} \right.  \\
    \mathrm{s.t.} ~&~ |\mathrm{Re}(\gamma_{m})| \leq \gamma_{\text{max}}, |\mathrm{Im}(\gamma_{m})| \leq \gamma_{\text{max}}, ~  \forall m.  
 \end{align}  
\end{subequations}
We propose to solve ${\bf Z}$ and ${\bm \gamma}$ in an iterative fashion as follows. 

\subsubsection{Update of ${\bf Z}$}
Firstly, we solve ${\bf Z}$ with a fixed ${\bm \gamma}$, i.e.,  \vspace{-1mm}
\begin{align}
\label{Problem_Z}
    \min_{{\bf Z}} ~ \frac{1}{2}\|{\bf Y} \! -  \left( {\bf I} + \text{diag}\{{\bm \gamma}\} \right){\bf Z}\|_{\text{F}}^{2} \! + \! \lambda_{1} \| [{\bf Z}, \mu {\bf I}] \|_{*}.
 \end{align}  
The derivative of the objective function in (\ref{Problem_Z}) with respect to ${\bf Z}$ is given as: $-{\bf D}^{\textrm{H}}{\bf Y} + {\bf D}^{\textrm{H}}{\bf D}{\bf Z} + \lambda_{1}{\bf P}{\bf Z}$, where ${\bf D} \triangleq {\bf I} + \text{diag}\{{\bm \gamma}\}$ and ${\bf P} \triangleq \left({\bf Z}{\bf Z}^{\textrm{H}} + \mu^{2} {\bf I} \right)^{-\frac{1}{2}}$. By setting the derivative to zero, we obtain  \vspace{-1mm}
\begin{align}
\label{Z_solution}
    {\bf Z} = \left( {\bf D}^{\textrm{H}}{\bf D} + \lambda_{1} {\bf P} \right)^{-1} {\bf D}^{\textrm{H}}{\bf Y}.
\end{align}

\subsubsection{Update of ${\bm \gamma}$}
Then, we solve ${\bm \gamma}$ with a fixed ${\bf Z}$, i.e.,  \vspace{-1mm}
\begin{subequations}
    \label{Problem_gamma}
    \begin{align}
      \min_{{\bm \gamma}} ~ & \frac{1}{2}\|{\bf g} - {\bm \Phi}{\bm \gamma} \|_{\text{2}}^{2} + \lambda_{2}(\|\mathrm{Re}({\bm \gamma})\|_{1} + \|\mathrm{Im}({\bm \gamma})\|_{1}), \\
      \mathrm{s.t.} ~&~ |\mathrm{Re}(\gamma_{m})| \leq \gamma_{\text{max}}, |\mathrm{Im}(\gamma_{m})| \leq \gamma_{\text{max}}, ~  \forall m,  
    \end{align} 
\end{subequations}
with $\|\cdot\|_{2}$ being the $\ell_{2}$ norm, where we have used the equation: $\text{vec}\{{\bf ADB}\} = ({\bf B}^{\textrm{T}} \odot {\bf A}){\bf d}$ with ${\bf D} = \text{diag}\{{\bf d}\}$ and $\odot$ being the Khatri-Rao product. In addition, in (\ref{Problem_gamma}), we define ${\bf g} \triangleq \text{vec}\{{\bf Y} \! - \! {\bf Z}\}$ and ${\bm \Phi} \triangleq {\bf Z}^{\textrm{T}} \odot {\bf I}$, with $\text{vec}\{\cdot\}$ being the vectorization opertation. 

Since $ {\bf g} = \mathrm{Re}({\bf g}) + \jmath \mathrm{Im}({\bf g})$ with $\jmath = \sqrt{-1}$, and similar for ${\bm \Phi}$ and ${\bm \gamma}$, we rewrite (\ref{Problem_gamma}) as   \vspace{-1mm}
 \begin{subequations}
\begin{align}
    \min_{{\bm \gamma}} ~ & \left\{ \!\!\!
    \begin{array}{l}
        \frac{1}{2}\| \mathrm{Re}({\bf g})- \mathrm{Re}({\bm \Phi})\mathrm{Re}(\bm \gamma)+\mathrm{Im}({\bm \Phi})\mathrm{Im}(\bm \gamma) \|_{\text{2}}^{2} \vspace{2mm} \\
        + \frac{1}{2}\| \mathrm{Im}({\bf g})- \mathrm{Re}({\bm \Phi})\mathrm{Im}(\bm \gamma)-\mathrm{Im}({\bm \Phi})\mathrm{Re}(\bm \gamma) \|_{\text{2}}^{2} \vspace{2mm} \\
        + \lambda_{2}(\|\mathrm{Re}({\bm \gamma})\|_{1}+\|\mathrm{Im}({\bm \gamma})\|_{1}),
    \end{array}  \right.  \\
 \mathrm{s.t.} ~&~ |\mathrm{Re}(\gamma_{m})|\leq \gamma_{\text{max}}, |\mathrm{Im}(\gamma_{m})|  \leq \gamma_{\text{max}}, ~ \forall m.
 \end{align}
 \end{subequations}
The above problem is a linearly constrained LASSO problem,  \vspace{-1mm}
\begin{align}
\label{reallasso}
    \min_{\bar{\bm \gamma}} ~ \frac{1}{2}\|\bar{\bf g} - \bar{\bm \Phi}\bar{\bm \gamma} \|_{\text{2}}^{2} + \lambda_{2}\|\bar{\bm \gamma}\|_{1}, ~~ \mathrm{s.t.} ~ \left\{ \!\! \begin{array}{l}
    \bar{\bm \gamma} \leq \gamma_{\text{max}}{\bf 1},  \vspace{1mm} \\
    -\bar{\bm \gamma} \leq \gamma_{\text{max}}{\bf 1},
    \end{array}
    \right.
\end{align}
where $\bar{\bf g} \!=\! [\mathrm{Re}({\bf g}) ; \mathrm{Im}({\bf g})]$, $\bar{\bm \Phi} \!=\! [\mathrm{Re}({\bm \Phi}),-\mathrm{Im}({\bm \Phi});\mathrm{Im}({\bm \Phi}),\mathrm{Re}({\bm \Phi})]$, $\bar{\bm \gamma} = [\mathrm{Re}({\bm \gamma}) ; \mathrm{Im}({\bm \gamma})]$, and ${\bf 1}$ is an all-one vector. Note that quadratic programming (QP) is a simple and efficient method for solving constrained LASSO problems \cite{Gaines2018}. We proceed to solve (\ref{reallasso}) by formulating it as a QP \cite{Gaines2018}: Let us first split $\bar{\bm \gamma}$ into positive and negative parts as $ \bar{\bm \gamma} = \bar{\bm \gamma}^+ - \bar{\bm \gamma}^-$. If $\bar{\gamma}_m \ge 0 $, then $\bar{\gamma}^+_i = \bar{\gamma}_i,\bar{\gamma}^-_i = 0$; else, $\bar{\gamma}^-_i = |\bar{\gamma}_i|,\bar{\gamma}^+_i = 0$. Then, Problem (\ref{reallasso}) becomes  \vspace{-1mm}
\begin{subequations}
\label{reallasso:qp}
\begin{align}
    \min_{\bar{\bm \gamma}^+,\bar{\bm \gamma}^-}& ~ \frac{1}{2}\|\bar{\bf g} - \bar{\bm \Phi}(\bar{\bm \gamma}^+-\bar{\bm \gamma}^-) \|_{\text{2}}^{2} + \lambda_{2} {\bf 1}^{\textrm{T}} (\bar{\bm \gamma}^+ + \bar{\bm \gamma}^-), \\ \mathrm{s.t.} ~&~ \left\{ \!\! \begin{array}{l}
    \bar{\bm \gamma}^+ \leq \gamma_{\text{max}}{\bf 1}, ~ \bar{\bm \gamma}^+ \ge {\bf 0}, \vspace{1mm} \\
    \bar{\bm \gamma}^- \leq \gamma_{\text{max}}{\bf 1}, ~ \bar{\bm \gamma}^- \ge {\bf 0}.
    \end{array}
    \right.
\end{align}
\end{subequations}
Problem (\ref{reallasso:qp}) can be efficiently solved by the existing QP solvers, such as \texttt{quadprog} function in MATLAB \cite{Optimization_Toolbox}.

\subsubsection{Overall algorithm for solving ${\bf Z}$ and ${\bm \gamma}$}
By iteratively updating ${\bf Z}$ and ${\bm \gamma}$ using (\ref{Z_solution}) and (\ref{reallasso:qp}), respectively, until convergence, we solve Problem (\ref{problem_nuclearl1_new}). The proposed algorithm is summarized in Algorithm \ref{Proposed_Alg}, where subscript $\cdot_{(\!k\!)}$ stands for the corresponding variable in the $k$-th iteration, $\epsilon$ is a small scalar and $k_{\text{max}}$ is a large scalar, used to terminate the algorithm. Besides, in Line 6 of Algorithm \ref{Proposed_Alg}, we define $f({\bf Z} , {\bm \gamma}) \triangleq \frac{1}{2}\|{\bf Y} \! - \! {\bf Z} \! - \! \text{diag}\{{\bm \gamma}\}{\bf Z}\|_{\text{F}}^{2} \! + \! \lambda_{1} \| [{\bf Z}, \mu {\bf I}] \|_{*} \! + \! \lambda_{2}\|{\bm \gamma}\|_{1}$.

\begin{algorithm}[t]
	\caption{Proposed algorithm for solving Problem (\ref{problem_nuclearl1_new})}
	\label{Proposed_Alg}
	\textbf{Input~~~~\!:} ${\bf Y} \in \mathbb{C}^{M \times T}$, $\lambda_{1}$, $\lambda_{2}$, $\mu$, $\epsilon$, $k_{\text{max}}$ \\
	\textbf{Output~~\!:} ${\bf{\widehat Z}} \in \mathbb{C}^{M \times T}$, ${\bm{\widehat \gamma}} \in \mathbb{C}^{M}$ \\
	\textbf{Initialize:} ${\bf Z}_{(\!0\!)} \gets {\bf Z}_{\text{init}}$, ${\bm \gamma}_{(\!0\!)} \gets {\bm \gamma}_{\text{init}}$, $k \gets 0$
	\begin{algorithmic}[1]
		\While {$k < k_{\text{max}}$}
        \State compute ${\bf D} \!=\! {\bf I} + \text{diag}\{ \! {\bm \gamma}_{(\!k\!)} \!\}$ and ${\bf P} \!=\! ({\bf Z}_{(\!k\!)}{\bf Z}_{(\!k\!)}^{\textrm{H}} + \mu^{2}{\bf I})^{-\frac{1}{2}}$
        \State update ${\bf Z}_{(\!k+1\!)}$ as ${\bf Z}_{(\!k+1\!)} = \left( {\bf D}^{\textrm{H}}{\bf D} + \lambda_{1} {\bf P} \right)^{-1} {\bf D}^{\textrm{H}}{\bf Y}$
        \State update ${\bm \gamma}_{(\!k+1\!)}$ by solving (\ref{reallasso:qp})
		\State $k \gets k+1$
		\State break $\gets$ $\frac{ | f({\bf Z}_{(\!k\!)} , {\bm \gamma}_{(\!k\!)}) - f({\bf Z}_{(\!k-1\!)} , {\bm \gamma}_{(\!k-1\!)}) | }{ | f({\bf Z}_{(\!k\!)} , {\bm \gamma}_{(\!k\!)}) | } \leq \epsilon $
		\EndWhile \State \textbf{end while} \\
		${\bf{\widehat Z}} \gets {\bf Z}_{(\!k\!)}$, ${\bm{\widehat \gamma}} \gets {\bm \gamma}_{(\!k\!)}$
	\end{algorithmic}
\end{algorithm}

It is worth highlighting the differences between the proposed method and our previous work \cite{Huang2023} as follows. \textit{(i)} In \cite{Huang2023}, we used a row-sparse matrix, say ${\bf V}$, to interpret the information introduced by the distorted sensors, and this matrix is treated independently from the noise-free matrix ${\bf Z}$. This results in loss of making use of the inherent relation between ${\bf Z}$ and ${\bf V}$. Differently, in this paper, we formulate the row-sparsity information as ${\bf V} = \text{diag}\{{\bm \gamma}\}{\bf Z}$, and thus we have taken into account the inherent relation between ${\bf Z}$ and ${\bf V}$. \textit{(ii)} In contrast to \cite{Huang2023}, we have considered the prior information on the maximal distortion level of the sensors in this work.

\subsubsection{Selection of hyper-parameters}
Note that $\mu$ should be as small as possible in order not to change the original problem, i.e., (\ref{problem_nuclearl1}). One efficient method is that we start with a proper value for $\mu$ and decrease its value at each iteration via $\mu_{(\!k+1\!)} = \alpha \mu_{(\!k\!)}$, with $\alpha < 1$. In our simulations, we start with $\mu_{(\!0\!)} = 1$ and $\alpha = 0.95$.

The tuning parameters $\lambda_{1}$ and $\lambda_{2}$ can be determined by applying, e.g., the cross-validation (CV) method \cite{Wu2020}, on (\ref{Problem_Z}) and (\ref{Problem_gamma}), respectively. However, since we have to apply the CV once per iteration, this will result in a super high computational complexity. Instead, we adopt the two-dimensional search strategy as introduced in \cite{Huang2023} (details can be found in \cite{Huang2023} and are omitted for brevity herein). In our simulations, we set $\lambda_{1} = 2$ and $\lambda_{2} = 0.2$.

\subsection{DOA estimation and distorted sensor detection}
\subsubsection{DOA estimation}
Denote the estimates of ${\bf Z}$ and ${\bm \gamma}$ as ${\widehat{\bf Z}}$ and $\widehat{\bm \gamma}$, respectively, and once they are obtained, they can be used for DOA estimation and distorted sensor detection. Note that ${\bf Z} = {\bf A}{\bf S}$ can be viewed as a noise-free data model. DOAs can be found via subspace-based methods, such as multiple signal classification (MUSIC), whose spatial spectrum is calculated as
$
P(\theta) = \frac{1}{{\bf a}^{\text{H}}(\theta)({\bf I} - {\bf L}{\bf L}^{\text{H}}){\bf a}(\theta)}
$ \cite{Schmidt1986}.
The singular value decomposition of $\widehat{\bf Z}$ is given by $\widehat{\bf Z} = {\bf L}{\bm \Sigma}{\bf R}^{\text{H}}$, where the columns of ${\bf L}$ and ${\bf R}$ contain the left and right orthogonal base vectors of $\widehat{\bf Z}$, respectively, and ${\bm \Sigma}$ is a diagonal matrix whose diagonal elements are the singular values of $\widehat{\bf Z}$ arranged in a descending order. We assume the number of sources, i.e., $K$, is known, and then the DOAs are determined by searching for the $K$ largest peaks of $P(\theta)$.

\subsubsection{Distorted sensor detection}
The number and positions of distorted sensors can be determined by the magnitude of the entries of $\widehat{\bm \gamma}$. 
Algorithm \ref{detec_failsensor} presents a strategy for detecting the distorted sensors. In words, we sort the modules of $\widehat{\bm \gamma}$ in an ascending order and obtain ${\bf{\tilde{\bm \gamma}}}$. We define the difference of the first two entries of ${\bf{\tilde{\bm \gamma}}}$ as $d = {\bf{\tilde{\bm \gamma}}}(2) - {\bf{\tilde{\bm \gamma}}}(1)$. Next, for $i = 3, 4, \cdots, M$, we compute ${\bf{\tilde{\bm \gamma}}}(i) - {\bf{\tilde{\bm \gamma}}}(i - 1)$ and compare it with a threshold, say $h$, of large value (we set $h = 10 d$ in our simulations in Section \ref{simulation}): if it is larger than or equal to $h$, we set $i_{\text{fail}} = i$ and break the for loop; if it is less than $h$, we have $i_{\text{fail}} = M + 1$. Finally, the number of distorted sensors is obtained as $M_{\text{fail}} = M - i_{\text{fail}} + 1$, and the corresponding sensors are the distorted sensors.

\begin{algorithm}[t]
	\caption{Detection of distorted sensors}
	\label{detec_failsensor}
	\textbf{Input~~:} ${\widehat{\bm \gamma}} \in \mathbb{C}^{M}$, $h$  \\
	\textbf{Output:} $M_{\text{fail}}$ \\
	~ calculate ${\bf{\tilde{\bm \gamma}}} = \text{sort}(|\widehat{\bm \gamma}| , \text{`ascend'})$ \\
	~ calculate $d = {\bf{\tilde{\bm \gamma}}}(2) - {\bf{\tilde{\bm \gamma}}}(1)$ and assign $i_{\text{fail}} = M + 1$
	\begin{algorithmic}[1]
		\For {$i = 3, 4, \cdots, M$}
		\If {${\bf{\tilde{\bm \gamma}}}(i) - {\bf{\tilde{\bm \gamma}}}(i - 1) \geq h$}
		\State $i_{\text{fail}} = i$  and break the \textbf{for} loop
		\EndIf \State \textbf{end if}
		\EndFor \State \textbf{end for} \\
		$M_{\text{fail}} \gets M - i_{\text{fail}} + 1$
	\end{algorithmic}
\end{algorithm}

\subsection{Complexity analysis}
The computational cost of the proposed Algorithm \ref{Proposed_Alg} mainly comes from the inverse operation for solving ${\bf Z}$ and the QP solver for solving ${\bm \gamma}$, which both are $\mathcal{O}(M^{3})$ \cite{Ye1989}. Therefore, the total computational cost of Algorithm \ref{Proposed_Alg} is $\mathcal{O}(N_{\text{iter}}M^{3})$, where $N_{\text{iter}}$ is the total number of iterations. Note that the cost of the IRLS method \cite{Huang2023} is $\mathcal{O}(N_{\text{irls}}M^{3})$, where $N_{\text{irls}}$ denotes the number of IRLS iterations.

\begin{figure}[t]  \vspace{-4mm}
\centering
{\includegraphics[width=0.5\textwidth]{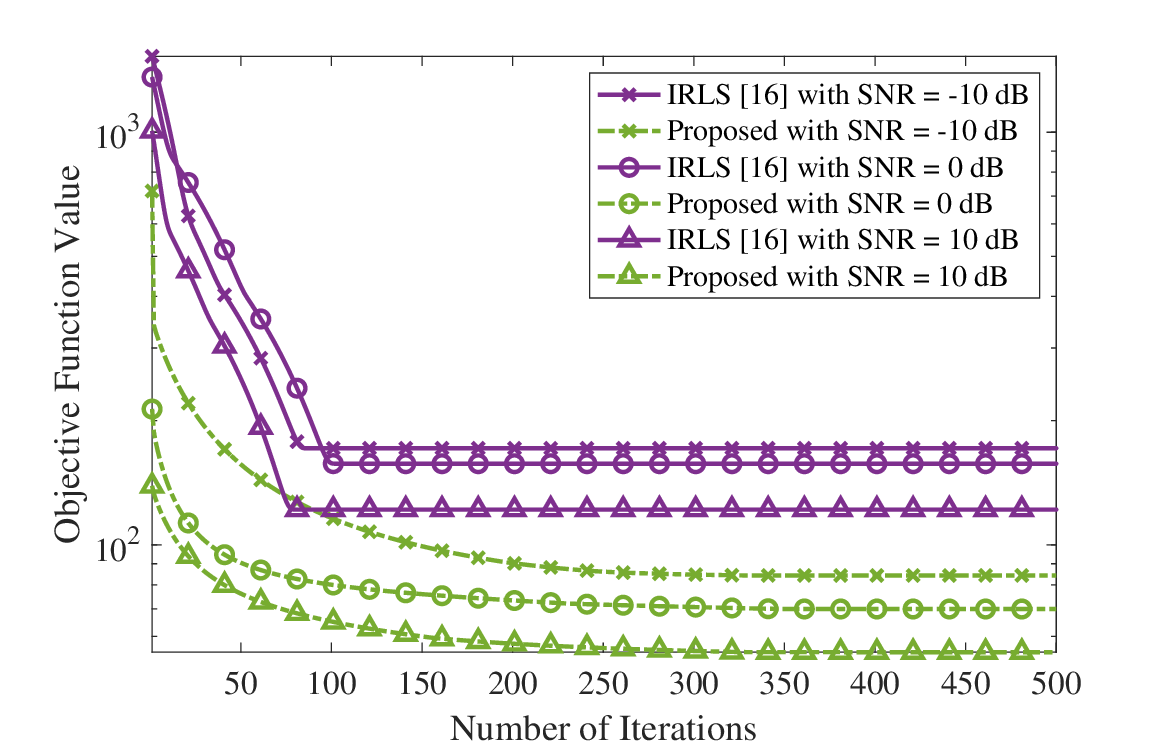}}
\caption{Convergence behavior of the proposed algorithm and the IRLS algorithm in \cite{Huang2023}.}
\label{ConvergenceBehavior}
\end{figure}

\begin{figure}[t]  \vspace{-4mm}
\centering
{\includegraphics[width=0.5\textwidth]{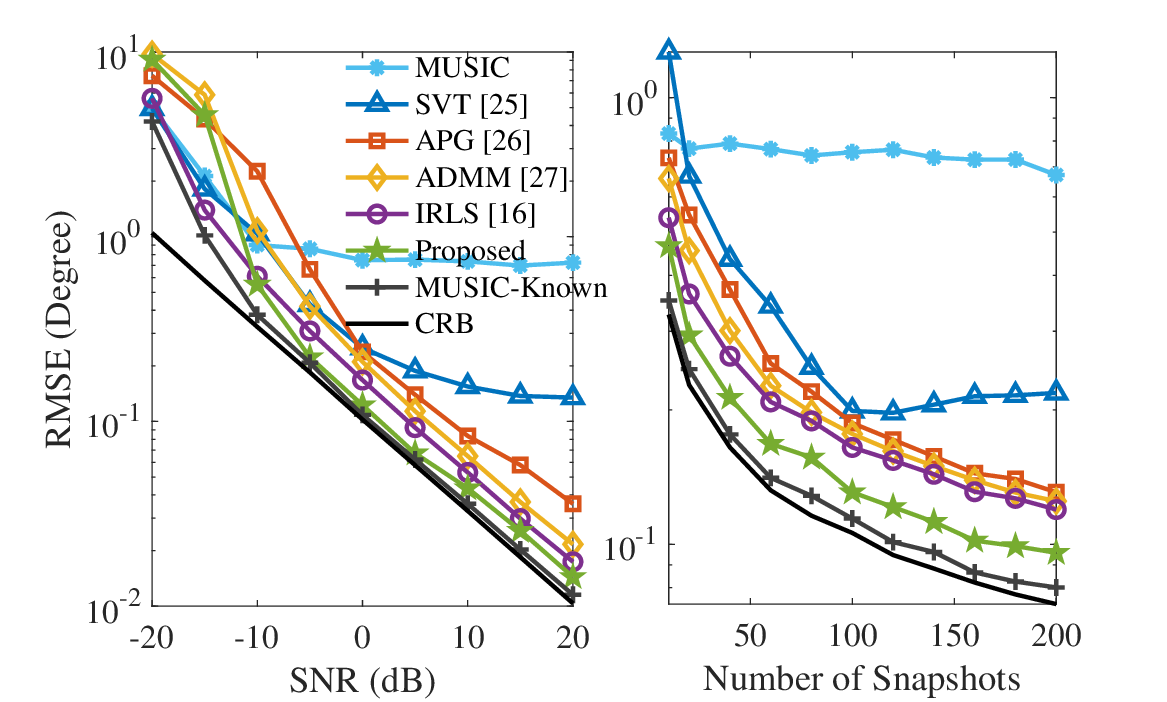}}
\caption{RMSE versus SNR (left) and number of snapshots (right).}   
\label{RMSEResults}
\end{figure}

\begin{figure}[t]  \vspace{-4mm}
\centering
{\includegraphics[width=0.5\textwidth]{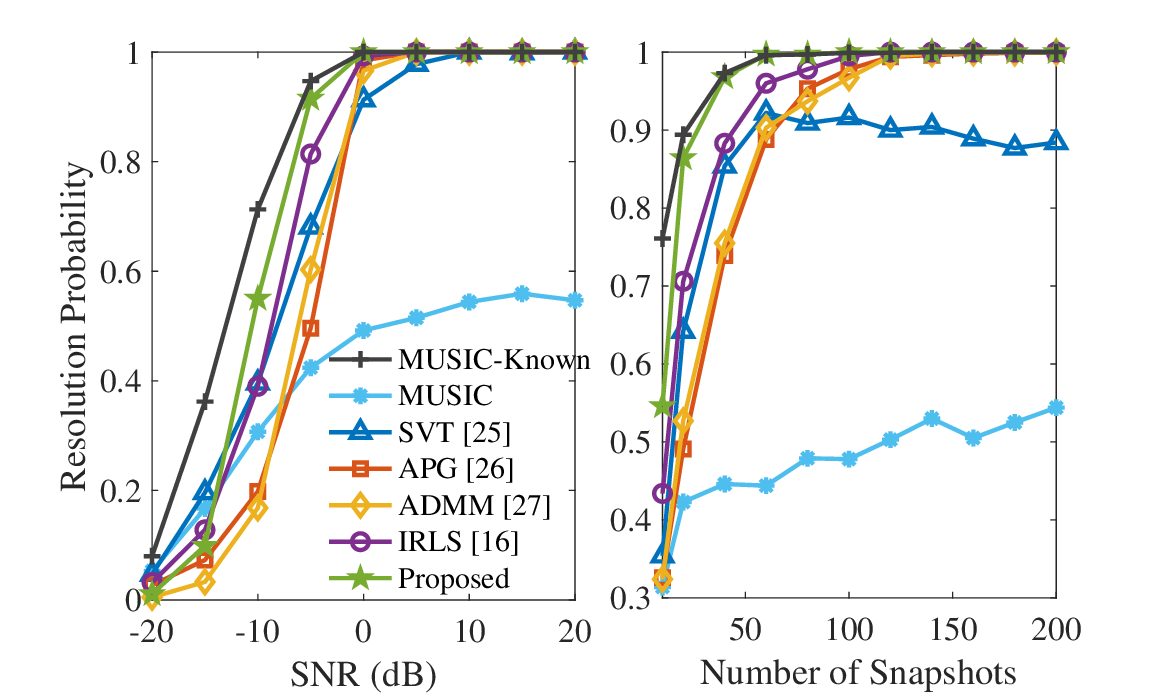}}
\caption{Resolution probability versus SNR (left) and number of snapshots (right).}   
\label{ResProbResults}
\end{figure}

\begin{figure}[t]  \vspace{-4mm}
\centering
{\includegraphics[width=0.5\textwidth]{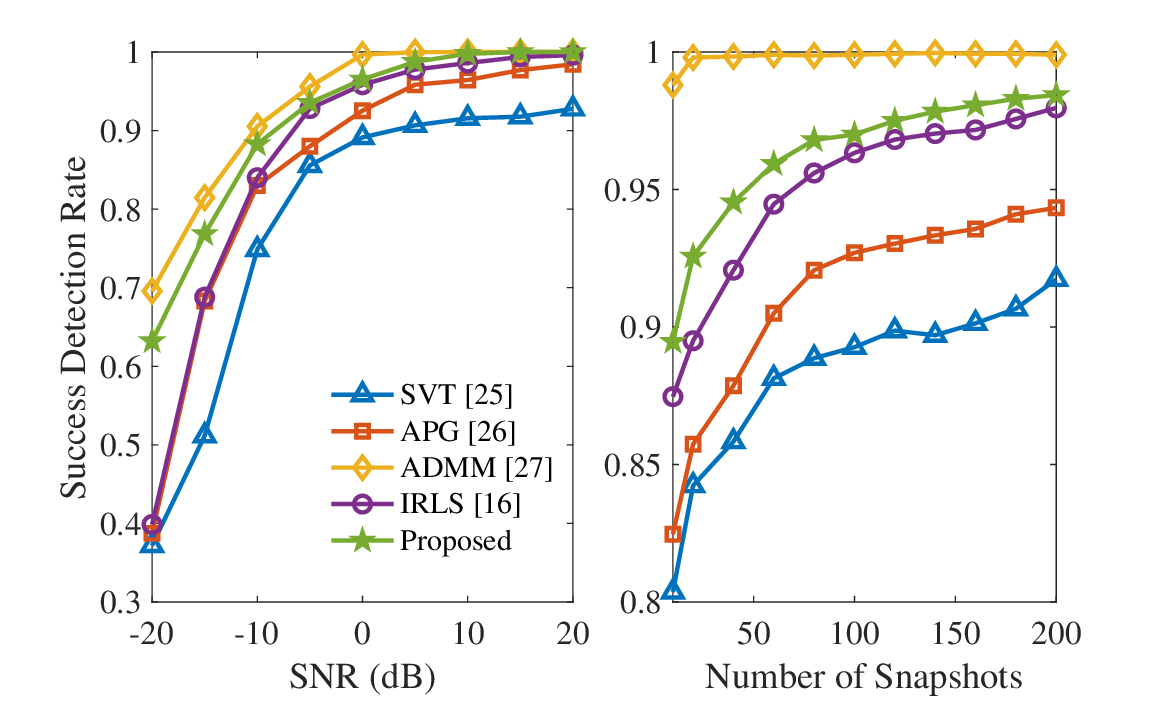}}
\caption{Success detection rate versus SNR (left) and number of snapshots (right).}   
\label{DetRateResults}
\end{figure}

\section{Simulations}
\label{simulation}

\subsection{Simulation scenario and performance indicators}   
We consider a ULA of $M = 8$ sensors, $3$ of which at random positions are distorted, $K = 2$ signals from $\pm 10^{\circ}$, signal-to-noise ratio $\text{SNR} = 10 \text{~dB}$, and $T = 100$ snapshots. The sensor gain and phase errors are randomly generated by drawing from uniform distributions on $[0, 10]$ and $[-10^{\circ}, 10^{\circ}]$, respectively.

The objective function values versus number of iterations are used as a metric for convergence behavior of the algorithms. Besides, we use the root-mean-squared-error (RMSE) and resolution probability (ResProb) to evaluate the DOA estimation performance, and the success detection rate (DetRate) to evaluate the performance of detecting distorted sensors. These three metrics are calculated via: $\text{RMSE} = \sqrt{\frac{1}{QK}\sum_{q = 1}^{Q}\sum_{k = 1}^{K}({\hat{\theta}}_{k,q} - \theta_{k})}$, $ \text{ResProb} = {N_{\text{succ}}}/{Q}$, and $\text{DetRate} = {N_{\text{detec}}}/{Q}$, where $Q$ is the number of Monte-Carlo runs, $\hat{\theta}_{k,q}$ denotes the DOA estimate of the $k$-th source in the $q$-th trial, $N_{\text{succ}}$ denotes the number of trials where $\max_{k} |\hat{\theta}_{k} - \theta_{k}| \leq 0.5^{\circ}$, and $N_{\text{detec}}$ is the number of trials where all distorted sensors are correctly found. In our simulations, $Q = 1000$, $\epsilon = 10^{-12}$, and $k_{\text{max}} = 100$.

\subsection{Simulation results and discussion}

\subsubsection{Convergence analysis}
First of all, we check the convergence behavior of the proposed algorithm and the IRLS algorithm \cite{Huang2023}. The objective function values versus number of iterations are plotted in Fig. \ref{ConvergenceBehavior}. It can be seen that both algorithms converge, and the IRLS converges faster than the proposed algorithm.

\subsubsection{Performance as a function of SNR}
The RMSE and ResProb of the methods versus SNR are depicted in Figs. \ref{RMSEResults} (left) and \ref{ResProbResults} (left), respectively, with $T = 100$. The traditional Cramér–Rao bound (CRB) with known sensor errors \cite{Delmas2014} is plotted as a benchmark. Note that the curve labelled as ``MUSIC-Known'' is the MUSIC with exact knowledge of the distorted sensors. It is seen that the singular value thresholding (SVT) \cite{Cai2010} and MUSIC have bad performance even when the SNR becomes large. The accelerated proximal gradient (APG) \cite{Beck2009}, alternating direction method of multipliers (ADMM) \cite{Lin2013}, IRLS \cite{Huang2023}, and the proposed algorithm perform well. When the SNR increases, their RMSEs decrease and their ResProbs increase up to $1$. Moreover, the proposed algorithm outperforms the other state-of-the-art methods in terms of RMSE and ResProb. The result of detecting distorted sensors is plotted in Fig. \ref{DetRateResults} (left), from which we see that the ADMM performs the best in terms of DetRate, followed by the proposed algorithm.

\subsubsection{Performance as a function of $T$}
The RMSE and ResProb of the methods versus number of snapshots are plotted in Figs. \ref{RMSEResults} (right) and \ref{ResProbResults} (right), respectively, with $\text{SNR} = 0 ~ \text{dB}$. The results demonstrate a better performance of the proposed algorithm compared with the SVT, APG, ADMM, and IRLS. On the other hand, the results of DetRate are presented in Fig. \ref{DetRateResults} (right). It is again seen that the ADMM has the best performance in terms of DetRate, followed by the proposed algorithm.

\section{Conclusion}
\label{conclusion}
The problem of joint direction-of-arrival (DOA) estimation and distorted sensor detection has been formulated under the framework of low-rank and row-sparse decomposition. The inherent connection between the low-rank and row-sparse components has been investigated and the maximal sensor distortion level has been taken into account. An alternating optimization method was proposed for solving the low-rank and sparse components, where a closed-form solution was derived for the low-rank component and a quadratic programming was developed for the sparse component. The proposed method was shown to outperform our previous work where the inherent connection was not explored. Simulation results demonstrated that the proposed method has excellent performance in DOA estimation and distorted sensor detection.

\balance
\bibliographystyle{myIEEEtran}       
\bibliography{refs}

\end{document}